\documentclass{Interspeech}

\interspeechcameraready

\title{Selective Invocation for Multilingual ASR: A Cost-effective Approach Adapting to Speech Recognition Difficulty}

\author[affiliation={1}]{Hongfei}{Xue}
\author[affiliation={2}]{Yufeng}{Tang}
\author[affiliation={2}]{Jun}{Zhang}
\author[affiliation={1}]{Xuelong}{Geng}
\author[affiliation={1}{*}]{Lei}{Xie}

\affiliation{Audio, Speech and Language Processing Group (ASLP@NPU), School of Software}{Northwestern Polytechnical University}{China}
\affiliation{}{ByteDance}{China}
\email{hfxue@mail.nwpu.edu.cn\thanks{*Corresponding author.}}

\usepackage{comment}
\usepackage{booktabs}
\usepackage{multirow}
\usepackage{hyperref}

\begin{document}

\maketitle

\begin{abstract}
Although multilingual automatic speech recognition (ASR) systems have significantly advanced, enabling a single model to handle multiple languages, inherent linguistic differences and data imbalances challenge SOTA performance across all languages. While language identification (LID) models can route speech to the appropriate ASR model, they incur high costs from invoking SOTA commercial models and suffer from inaccuracies due to misclassification. To overcome these, we propose SIMA, a selective invocation for multilingual ASR that adapts to the difficulty level of the input speech. Built on a spoken large language model (SLLM), SIMA evaluates whether the input is simple enough for direct transcription or requires the invocation of a SOTA ASR model. Our approach reduces word error rates by 18.7\% compared to the SLLM and halves invocation costs compared to LID-based methods. Tests on three datasets show that SIMA is a scalable, cost-effective solution for multilingual ASR applications.
\end{abstract}


\section{Introduction}
\label{sec:intro} 

\begin{figure}[t]
\centering
\includegraphics[width=1.0\linewidth]{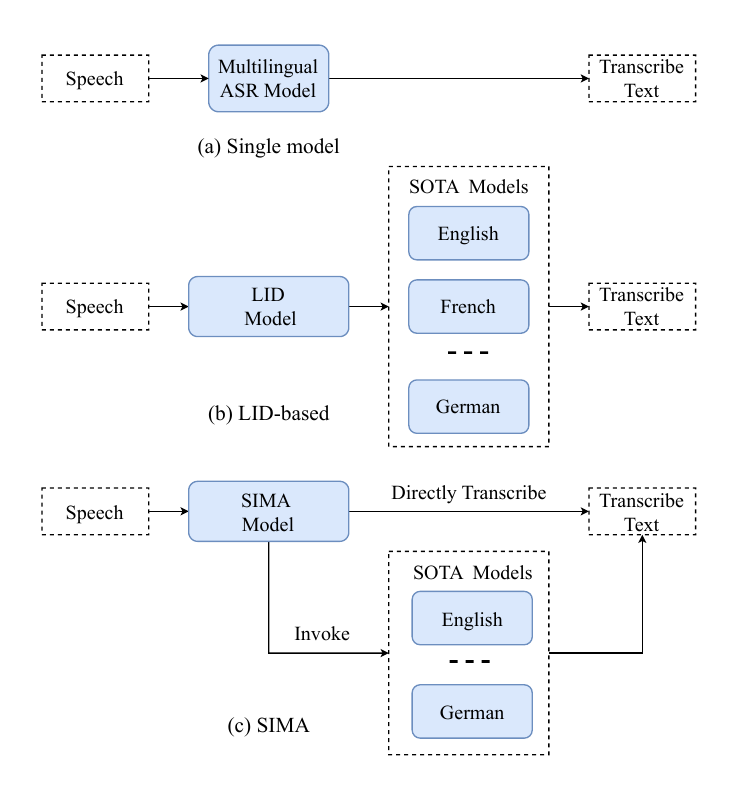}
\caption{Three systems for multilingual ASR. (a) A single multilingual model, such as Whisper, which recognizes multiple languages with one model. (b) A language identification (LID)-based system that identifies the language and invokes the corresponding SOTA model. (c) Selective invocation for multilingual ASR (SIMA) that directly transcribes simpler speech and invokes SOTA models for more complex inputs.}
\label{fig:model}
\vspace{-20pt}
\end{figure}

Multilingual automatic speech recognition (ASR) models have gained significant attention for their ability to recognize multiple languages using a single model~\cite{18multilingual, conneau2019cross, 21XLSR, 23mlsuperb}, as illustrated in Figure~\ref{fig:model}(a). Recent advances have led to impressive performance in various languages through large-scale supervised or self-supervised pre-training~\cite{21XLSR, 22JUST, 23whisper, xue2023tranusr, 23mms, 23usm, 23seamlessm4t, 24SSHR, universal-1}. For example, Whisper~\cite{23whisper} is trained on 680,000 hours of weakly multilingual data, enabling it to generalize effectively across standard ASR benchmarks, while USM~\cite{23usm} leverages 12 million hours of unlabeled data to achieve robust cross-lingual performance.
Despite these advances, the application of multilingual ASR systems with a single model still faces significant challenges. Phonetic differences, syntactic variations, and vocabulary disparities across languages make it difficult to achieve consistent universal state-of-the-art (SOTA) performance. Moreover, imbalances in training data between high-resource and low-resource languages further limit the single-model solutions.

A common strategy to address these challenges is to use a language identification (LID) model that first detects the language of the input speech before invoking the corresponding SOTA ASR model for transcription, as shown in Figure~\ref{fig:model}(b). 
However, this two-stage approach has its drawbacks. Many SOTA models are commercial~\cite{universal-1} and incur usage fees based on the volume of processing, making this method costly. Additionally, an incorrect LID prediction may trigger the wrong model, further affecting the user experience~\cite{houston2024improving}.

Motivated by these limitations, we propose an alternative strategy that selectively invokes models based on the complexity of the input speech. In ASR tasks, the recognition difficulty varies significantly. Under clean acoustic conditions with simple vocabulary, both the SOTA and regular models typically yield low word error rates (WER). However, in noisy or acoustically challenging environments, the WER increases~\cite{12noise, 15musan, 22wavlm, 23noiseasr}, where robust SOTA models generally perform better~\cite{23whisper}. This observation raises a key question: Can we distinguish between simple and complex speech inputs and adapt our ASR system accordingly? Recent advancements in large language models (LLMs)~\cite{LLaMA, bai2023qwen, openai2023gpt4} have significantly enhanced the understanding capabilities of spoken large language models (SLLMs)~\cite{23speechllama, chu2023qwenaudio, 24wavllm, chu2024qwen2audio, 24echat}. We hypothesize that SLLMs can not only understand the content of speech but also assess whether they can transcribe the speech accurately. In this way, the SLLM could directly transcribe simple, clean speech while reserving the invocation of a more robust SOTA model for complex cases.

Based on this hypothesis, we introduce \textbf{S}elective \textbf{I}nvocation for \textbf{M}ultilingual \textbf{A}SR (SIMA), built on a base SLLM. As illustrated in Figure~\ref{fig:model}(c), SIMA processes the input speech and determines whether to transcribe it directly or invoke a specialized SOTA model for more challenging inputs. Specifically, for simple speech or when the language confidence is low, SIMA generates the transcription itself, thereby avoiding the high costs associated with invoking commercial SOTA models. Furthermore, to more accurately determine whether an invocation is needed, we introduce an ``Uncertain" category as a supplement to the binary classification. For these ``Uncertain" cases, we employ a fusion confidence strategy that integrates three distinct confidence measures to assess.
Experiments on the Multilingual Librispeech~\cite{20mllibrispeech}, VoxPopuli~\cite{20voxpopuli}, and FLEURS~\cite{22FLEURS} datasets demonstrate that SIMA reduces the WER by 18.7\% relative to the base SLLM, while halving invocation costs compared to the LID-based approach. These results support our hypothesis and demonstrate that SIMA offers a scalable and cost-effective solution for multilingual ASR applications.



\section{Method}
\label{sec:proposed_method}
As illustrated in Figure~\ref{fig:model}(c), our proposed method comprises two main modules. The first module is the SIMA model, which evaluates the difficulty of the input speech and either directly produces a transcription or outputs an invocation label. The second module is a library of SOTA ASR models. When the SIMA model yields the invocation label, the input speech is routed to the appropriate SOTA model for transcription.

\subsection{SIMA Design}
\begin{figure}[t]
\centering
\includegraphics[width=1.0\linewidth]{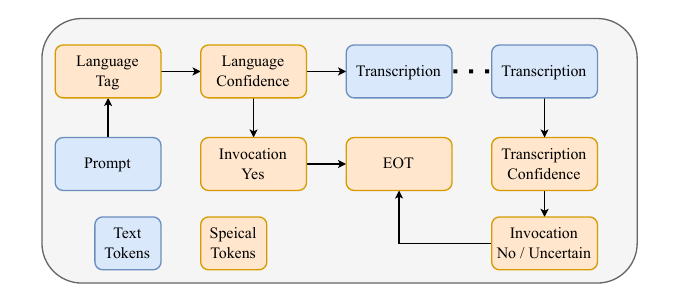}
\caption{The multitask training format of the SIMA model.}
\label{fig:task-format}
\end{figure}

The SIMA model is built upon a multilingual SLLM and is designed to produce one of three possible outputs during training, as depicted in Figure~\ref{fig:task-format}. Each output includes a predicted language tag along with an associated language confidence score. The three output types are defined as follows:

\noindent\textbf{Invocation No:} For speech deemed sufficiently simple, the model generates transcription tokens followed by an ``Invocation No" token, indicating that it can directly transcribe the input. During training, the ground truth transcription and special tokens are used to compute the cross-entropy (CE) loss.

\noindent\textbf{Invocation Yes:} For complex speech, the model outputs an ``Invocation Yes" token, signaling that a specialized SOTA model should be invoked to process the input. In this case, the CE loss is computed solely based on the special token.

\noindent\textbf{Invocation Uncertain:} We found that a binary decision (i.e., simply ``Yes" or ``No") can be ambiguous for speech of intermediate difficulty. To address this, we introduce a third category—``Invocation Uncertain"—which indicates uncertainty and the need for further evaluation via a confidence strategy. This output is formatted similarly to the ``No" output; however, during training, the transcription tokens are pseudo labels for training the ``Transcription Confidence" token, and the CE loss is only computed for special tokens.

It is important to note that when the language confidence score is low, the SIMA model defaults to direct transcription to avoid erroneous invocations.

\begin{figure}[t]
\centering
\includegraphics[width=1.0\linewidth]{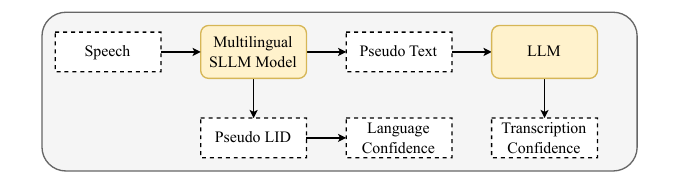}
\caption{Data pipeline of the SIMA dataset.}
\label{fig:data-pipline}
\vspace{-10pt}
\end{figure}

\subsection{Fusion Confidence Strategy}
To address cases classified as ``Invocation Uncertain" and determine whether a language-specific invocation is necessary, we integrate three complementary confidence evaluation methods:

\noindent\textbf{Posterior Probability:} We extract the maximum probability from the softmax output of the final layer, where $y \in \mathbb{R}^{T \times C}$, with $T$ denoting the number of frames and $C$ representing the number of label classes. The average posterior probability is computed as follows:
\begin{align}
\text{Probability} = \frac{1}{T}\sum_{t=1}^{T}\max_{c \in \{1, \ldots, C\}} y_{t,c}
\end{align}
\noindent\textbf{Entropy:} Since entropy is a well-known measure of uncertainty, we employ an entropy-derived confidence score defined by:
\begin{align}
\text{Entropy} = -\frac{1}{T \times C}\sum_{t=1}^{T}\sum_{c=1}^{C} y_{t,c}\log(y_{t,c})
\end{align}
\noindent\textbf{Transcription Confidence:} After generating transcription tokens, the SIMA model generates an evaluation score that reflects sentence fluency. This score categorizes the transcription into four confidence levels, ranging from level A (highest) to level D (lowest).

In the fusion strategy, if the posterior probability falls below a threshold $P$, the entropy exceeds a threshold $E$, and the transcription confidence is lower than a threshold $T$, the system determines that the speech recognition is difficult for direct transcription and thus invokes the appropriate SOTA model.

\subsection{Data Pipline} 
\label{sec: dataset}

To train the SIMA model, we propose a specialized data pipeline that supports the three distinct output formats. We employ a multilingual SLLM~\cite{xue2024ideal} as the base model for generating training data. As illustrated in Figure~\ref{fig:data-pipline}, the SLLM processes the input speech to produce pseudo text labels, from which we compute the corresponding WER.
Based on the WER, we assign invocation labels for all languages as follows:
\begin{itemize}
    \item \textbf{Invocation No:} WER in the interval [0, 2].
    \item \textbf{Invocation Yes:} WER exceeding 10.
    \item \textbf{Invocation Uncertain:} WER in the interval (2, 10].
\end{itemize}

Since SOTA model performance (i.e., measured by WER) can vary across languages, we further analyze the results at different WER intervals for each language in Section~\ref{sec: language-specific}.

After obtaining the transcription and invocation labels, we utilize an LLM~\cite{abdin2024phi} to score the generated text against the ground truth, thereby producing transcription confidence scores ranging from level A (highest) to level D (lowest). Additionally, the SLLM is used to infer a pseudo LID and its associated posterior probability. If the SLLM misidentifies the language, the transcription is assigned the lowest confidence level (D); otherwise, confidence levels from D to A are determined based on the posterior probability. This design enables the dynamic adjustment of invocation thresholds, reducing the likelihood of erroneous model invocations.

\begin{table}[t]
\centering
\caption{WER (\%) in LID-based models in MLS testset.}
\label{tab:SOTA-result}
\resizebox{\linewidth}{!}{
\begin{tabular}{@{}lcccccccc@{}}
\toprule
WER         & \textit{de}            & \textit{en}            & \textit{es}            & \textit{fr}            & \textit{it}            & \textit{nl}            & \textit{pl}            & Avg           \\ \midrule
OpenAI & 4.68          & 5.81          & 4.26          & 5.82          & 9.95          & 9.89          & \textbf{4.52} & 6.42          \\
Meta  & 4.83          & 7.34          & \textbf{4.02} & \textbf{5.45} & \textbf{9.62} & 12.11         & 7.00          & 7.20          \\
Assembly      & \textbf{3.91} & \textbf{5.23} & 4.73          & 7.26          & 13.69         & \textbf{9.81} & 5.57          & 7.71          \\
LID-Top             & \textbf{3.91} & \textbf{5.23} & \textbf{4.02} & \textbf{5.45} & \textbf{9.62} & \textbf{9.81} & \textbf{4.52} & \textbf{6.08} \\ \bottomrule
\end{tabular}
}
\vspace{-10pt}
\end{table}

\section{Experiments}
\label{sec:experiments}
\subsection{Datasets}
To ensure domain diversity and improve robustness, we utilize three datasets: Multilingual LibriSpeech (MLS)~\cite{20mllibrispeech}, VoxPopuli~\cite{20voxpopuli}, and FLEURS~\cite{22FLEURS}. We select the languages common to these datasets, namely English (\textit{en}), German (\textit{de}), Dutch (\textit{nl}), French (\textit{fr}), Spanish (\textit{es}), Italian (\textit{it}), and Polish (\textit{pl}). 
For training, we generate 100k samples per language using the SIMA data pipeline, with 25k samples sourced from VoxPopuli and 75k from MLS. This division is due to the size of the original dataset.
For validation and testing, SIMA data are generated from the original validation and test sets of MLS and VoxPopuli. For FLEURS, only the test set is used as out-of-domain data to evaluate robustness further.

\subsection{Experiment Setup}
\textbf{Base Model}  
The base model used for SIMA initialization comprises speech encoders, an adapter, and a text LLM, following the architecture described in IdealLLM~\cite{xue2024ideal}. We train the model on the MLS and VoxPopuli datasets to perform both ASR and LID tasks simultaneously, meeting the requirements of SIMA pipeline. It also serves as our baseline for comparison.

\noindent\textbf{SIMA Model}  
The SIMA model is initialized with the base model and fine-tuned on the generated SIMA data using a multi-task learning framework. During training, we maintain a balanced distribution among the three output classes in an approximate ratio of 1:1.5:1.5. 
For the fusion confidence strategy, we set $P$ to 0.96, $E$ to 0.0015, and $T$ to level B based on the multiple tests on the validation set.
Training is performed on 8 NVIDIA A100 GPUs, with gradient accumulation configured to achieve an effective batch size corresponding to 400 seconds of speech per GPU. We use a peak learning rate of 5e-5, a warmup period of 2k steps, and 10k training steps.

\noindent\textbf{Random Invocation Model}  
To isolate the effectiveness of our proposed method from improvements arising solely from the invocation mechanism, we design a random invocation baseline. In this baseline, the decision to invoke is made randomly at the same overall invocation rate as the SIMA model. Additionally, we propose an \textbf{invocation efficiency} metric to quantify the benefit of each invocation, defined as the reduction in WER relative to the base model divided by the invocation rate. The invocation rate is the proportion of input speech invoking the SOTA model.

\noindent\textbf{LID-based Model}  
To establish LID-based benchmarks, we evaluate some open-source models and commercial models on the test sets of MLS, VoxPopuli, and FLEURS. For each dataset, we select the model achieving the lowest WER as the SOTA reference for this study. 
The models considered for comparison included OpenAI's API\footnote{https://platform.openai.com/docs/guides/speech-to-text}, Meta's SeamlessM4T-large-v2\footnote{https://huggingface.co/facebook/seamless-m4t-v2-large}, and AssemblyAI's universal-1\footnote{https://www.assemblyai.com/research/universal-1}. 
For more commercial models, they are not used due to cost constraints.
The performance results for MLS are summarized in Table~\ref{tab:SOTA-result}, and the same procedure is applied to obtain SOTA results for VoxPopuli and FLEURS.
It should be noted that the error caused by the incorrect LID prediction is not considered in LID-Top. Considering the SOTA model uses multilingual models, the impact is not significant.

\begin{table}[t]
\centering
\caption{WER (\%) $\downarrow$ results and Invocation Rate(\%) $\downarrow$ results.}
\label{tab:mainresult}
\resizebox{\linewidth}{!}{
\begin{tabular}{@{}lcccccc@{}}
\toprule
           & \multicolumn{2}{c}{MLS} & \multicolumn{2}{c}{VoxPopuli} & \multicolumn{2}{c}{FLEURS} \\ \midrule
           & WER      & Rate     & WER       & Rate     & WER         & Rate      \\ \midrule
Whisper    & 6.42      & 0           & 14.00     & 0           & 5.06        & 0            \\
LID-Top       & 6.08      & 100         & 11.01     & 100         & 5.06        & 100          \\
Base & 7.86      & 0           & 12.43     & 0           & 10.76       & 0            \\
Random   & 6.85      & 57.6       & 11.78     & 45.5       & 7.73        & 51.2        \\
SIMA         & 6.40      & 57.6       & 11.28     & 45.5       & 6.43        & 51.2        \\ \bottomrule
\end{tabular}
}
\end{table}

\begin{table}[t]
\centering
\caption{Others' results of SIMA model and LID-based model.}
\label{tab:mainresult2}
\resizebox{\linewidth}{!}{
\begin{tabular}{@{}lcccc@{}}
\toprule
             & \multicolumn{1}{c}{MLS} & \multicolumn{1}{c}{Vox} & \multicolumn{1}{c}{FLEURS} & \multicolumn{1}{c}{Avg}\\ \midrule
SIMA-ACC (\%)     & 72.1                    & 69.7                    & 72.8     & 71.5                  \\
SIMA-F1 (\%)           & 73.6                    & 68.1                    & 68.1     & 69.9                  \\
SIMA-Cost (×)         & 0.58                    & 0.46                    & 0.51      & 0.51                 \\
SIMA-Invoke-Errors (\%) & 0.12                    & 0.74                    & 0.18     & 0.35                  \\
LID-Invoke-Errors (\%) & 0.43                    & 1.22                    & 0.38     & 0.68                  \\
\bottomrule
\end{tabular}
}
\vspace{-10pt}
\end{table}

\begin{table*}[h]
\centering
\caption{Results of Language-Agnostic and Language-Specific in the MLS test set.}
\label{tab:language}
\begin{tabular}{@{}llcccccccc@{}}
\toprule
\textbf{}                                 &            & \textit{de}    & \textit{en}    & \textit{nl}    & \textit{es}    & \textit{fr}    & \textit{pl}    & \textit{it}    & \multicolumn{1}{l}{Avg} \\ \midrule
\multirow{5}{*}{Language-Agnostic} & ACC $\uparrow$        & 72.75 & 69.01 & 80.85 & 76.02 & 66.24 & 65.96 & 73.53 & \textbf{72.05}                   \\
\multicolumn{1}{c}{}                      & F1 $\uparrow$         & 69.01 & 74.56 & 88.6  & 68.61 & 64.19 & 71.41 & 78.71 & \textbf{73.58}                   \\
\multicolumn{1}{c}{}                      & Rate $\downarrow$       & 49.12 & 72.3  & 87.71 & 38.57 & 53.17 & 47.31 & 55.15 & 57.62                   \\
\multicolumn{1}{c}{}                      & WER $\downarrow$        & 3.95  & 5.23  & 9.65  & 4.27  & 5.26  & 6.29  & 10.21 & \textbf{6.41}                    \\
\multicolumn{1}{c}{}                      & Efficiency $\uparrow$ & 2.28  & 1.55  & 2.60  & 1.63  & 0.51  & 9.07  & 0.80  & 2.52                    \\ \midrule
\multirow{5}{*}{Language-Specific}                   & ACC $\uparrow$        & 72.01 & 69.83 & 68.75 & 72.79 & 68.88 & 57.5  & 73.06 & 68.97                   \\
  & F1 $\uparrow$         & 68.58 & 73.9  & 70.26 & 63.47 & 60.69 & 61.7  & 68.35 & 66.70                   \\
  & Rate $\downarrow$       & 41.46 & 67.37 & 51.32 & 30.69 & 41.38 & 37.31 & 36.13 & \textbf{43.66}                   \\
  & WER $\downarrow$        & 4.03  & 5.24  & 9.87  & 4.38  & 5.25  & 7.04  & 10.36 & 6.59                    \\
  & Efficiency $\uparrow$ & 2.53  & 1.65  & 4.01  & 1.69  & 0.68  & 9.49  & 0.80  & \textbf{2.90}                    \\ \bottomrule
\end{tabular}
\vspace{-10pt}
\end{table*}

\subsection{Main Results}
Table~\ref{tab:mainresult} summarizes the performance of the SIMA and baseline models regarding WER and invocation rate across the three test sets. The results indicate that, due to the selective invocation of SOTA models, the SIMA model achieves significant WER reductions of 18.6\%, 9.3\%, and 28.2\% relative to the base model on the three datasets.
Furthermore, compared to the random invocation strategy, SIMA consistently delivers lower WER, with improvements of 6.6\%, 4.2\%, and 16.8\%. Notably, the improvement on the FLEURS dataset is especially significant, as it is out-of-domain for the base model but in-domain for the LID-Top model. These findings convincingly demonstrate SIMA’s remarkable ability to precisely determine when to invoke the SOTA model, thereby optimizing overall ASR performance.

Table~\ref{tab:mainresult2} presents additional performance metrics for SIMA. The invocation decision accuracy (ACC) and F1 scores are approximately 70\%, supporting our hypothesis that SLLMs can effectively differentiate speech inputs based on complexity. Although SIMA exhibits a slight WER gap compared to LID-Top, it reduces invocation costs by approximately 0.51× across the three datasets, significantly lowering associated expenses. Moreover, incorporating language confidence prediction reduces language invocation errors (SIMA-Invoke-Errors) by about 48.6\% relative to LID-based methods, with even greater impact when the SOTA model is monolingual.

\subsection{Language-Specific Invocation Strategies}
\label{sec: language-specific}
Due to the varying recognition performance (i.e., WER) across different languages in both SOTA and base models, a unified invocation strategy may lead to unnecessary model invocations for some languages. Table~\ref{tab:language} compares the unified (language-agnostic) strategy with language-specific strategies. As introduced in Section~\ref{sec: dataset}, the language-agnostic strategy uses a fixed invocation interval. In contrast, the language-specific strategy customizes this interval: for a given language with a LID-Top model test result of $i$, the ``Uncertain" interval is defined as ($i-2.5$, $i+2.5$]. This tailored approach enhances the efficiency of invoking the SOTA model by adapting to language-specific characteristics.

The results in Table~\ref{tab:language} indicate that while language-specific strategies slightly decrease the ACC and F1 scores of invocation decisions (reflecting the increased complexity of the model's learning task), they also significantly reduce the overall invocation rate, improving the invocation efficiency metric from 2.5 to 2.9. These findings suggest that language-specific invocation strategies can optimize the overall efficiency of the SIMA model by minimizing unnecessary invocations without substantially compromising WER.

\subsection{Analysis}

\textbf{Ablation Study}
Table~\ref{tab:ablation} presents the results of ablation study on MLS, which focuses on two key components: fusion confidence strategy and the use of ``Invocation Uncertain". First, we remove the fusion confidence strategy and replace it with random invocations, forcing all predictions marked as ``Uncertain" to be classified definitively as either ``Yes" or ``No." This change led to an increased WER, alongside declines in ACC, F1 scores, and efficiency. These findings highlight the confidence strategy's crucial role in enhancing the precision of model invocations.
In the second experiment, we eliminate the ``Uncertain" category, compelling the system to make a binary decision. This binary approach increased the overall invocation rate and reduced invocation efficiency, suggesting that it triggers unnecessary invocations for audio samples that are simple and clean. Moreover, the observed decrease in ACC indicates that samples with intermediate difficulty are more susceptible to misclassification when the ``Uncertain" option is unavailable.

\begin{table}[t]
\centering
\caption{Ablation study on confidence strategy and uncertainty.}
\resizebox{1.0\linewidth}{!}{
\label{tab:ablation}
\begin{tabular}{@{}lccccc@{}}
\toprule
            & ACC $\uparrow$   & F1 $\uparrow$   & Rate $\downarrow$ & WER $\downarrow$  & Efficiency $\uparrow$  \\ \midrule
SIMA         & \textbf{72.1}  & \textbf{73.6} & \textbf{57.6}   & 6.40 & \textbf{2.52}       \\
-- Confidence & 69.6  & 71.5 & 57.6  & 6.47  & 2.24       \\
-- Uncertain  & 67.9 & 71.3 & 66.0 & \textbf{6.39}   & 2.21       \\ \bottomrule
\end{tabular}
}
\vspace{-5pt}
\end{table}

\noindent\textbf{ASR Performance of SIMA}
Table~\ref{tab:asr_perf} compares the ASR performance of SIMA against the base SLLM. Since SIMA does not generate transcription results for samples that invoke the SOTA model, we evaluate its performance on subsets from the MLS, Vox, and FLEURS datasets—comprising only those samples transcribed directly by SIMA. The results reveal that SIMA training does not increase the WER; in fact, the WER slightly decreases in some cases. This outcome may be attributed to the evaluated subset primarily consisting of simple and clean speech samples that benefit from further training optimization. In contrast, more complex samples (routed to the SOTA model) are excluded from the training loss calculation.

\begin{table}[]
\centering
\caption{WER (\%) results of SIMA's direct transcription.}
\label{tab:asr_perf}
\resizebox{0.95\linewidth}{!}{
\begin{tabular}{@{}lcccc@{}}
\toprule
     & MLS  & VoxPopuli   & FLEURS & Avg  \\ \midrule
Base SLLM & 7.81 & 10.79 & 10.33  & 9.64 \\
SIMA  & 7.79 & 9.72  & 10.39  & 9.30 \\ \bottomrule
\end{tabular}
}
\vspace{-10pt}
\end{table}

\noindent\textbf{Future Work}
Although the current SIMA model significantly improves WER, it still lags behind Whisper~\cite{23whisper} on out-of-domain data, FLEURS~\cite{22FLEURS}. This limitation stems from our initial hypothesis that the base SLLM model can effectively perform the invoke task. Our base SLLM model~\cite{xue2024ideal} is inherently weaker than specialized models such as Whisper because of the limitation of training data. In future work, we plan to adopt Whisper~\cite{23whisper} as the base model and further refine the SIMA system to improve the ASR performance of the SOTA model. 

\section{Conclusion}
\label{sec:conclusion}
This paper introduces SIMA, a novel selective invocation strategy for multilingual ASR. Leveraging a base spoken large language model, SIMA dynamically determines whether to transcribe speech directly or invoke specialized SOTA models. Extensive experiments on three benchmark datasets demonstrate that SIMA reduces the word error rate by 18.7\% and cuts invocation costs by 51\% compared to LID-based methods. These promising results highlight the potential of adaptive ASR systems for scalable, cost-effective real-world applications. In future work, we will explore stronger base models to further enhance performance.


\newpage

\bibliographystyle{IEEEtran}
\bibliography{mybib, refs}

\end{document}